\documentclass[journal,comsoc]{IEEEtran}

\usepackage[T1]{fontenc}
\usepackage{cite}
\usepackage{amsmath}
\usepackage{subcaption}
\usepackage[cmintegrals]{newtxmath}
\usepackage[hidelinks]{hyperref}
\usepackage{graphicx}
\usepackage{multirow}
\usepackage{qcircuit}
\usepackage{orcidlink}

\usepackage{tikz}
\usetikzlibrary{shapes.geometric, arrows}

\tikzstyle{startstop} = [rectangle, rounded corners, minimum width=1cm, minimum height=1cm,text centered, draw=black, fill=red!10]

\tikzstyle{io} = [trapezium, trapezium left angle=70, trapezium right angle=110, minimum height=1cm, text centered, draw=black, fill=blue!5]

\tikzstyle{process} = [rectangle, minimum width=3cm, minimum height=1cm, text centered, draw=black, fill=orange!10]

\tikzstyle{decision} = [diamond, minimum width=3cm, minimum height=1cm, aspect=2.2, text centered, draw=black, fill=green!10]

\tikzstyle{arrow} = [thick,->,>=stealth]

\newcommand{\ket}[1]{| #1 \rangle}
\newcommand{\bra}[1]{\langle #1 |}
\newcommand{\kbra}[2]{|#1\rangle\!\langle #2|}
\newcommand{\bket}[2]{\langle #1|#2\rangle}
\DeclareMathOperator{\tr}{tr}

\def\papertype{article}

\usepackage{xcolor}
\definecolor{amber}{rgb}{1.0, 0.75, 0.0}

\usepackage{etoolbox}
\robustify\cite

\usepackage[final]{changes}
\def\addedStart{  }
\def\addedEnd{ }

\begin{document}
\title{Variational Quantum-Based Simulation of Waveguide Modes}

\author{
    Wei-Bin~Ewe\orcidlink{0000-0002-4600-0634},
    Dax~Enshan~Koh\orcidlink{0000-0002-8968-591X}, Siong~Thye~Goh\orcidlink{0000-0001-7563-0961}, Hong-Son~Chu\orcidlink{0000-0002-7077-0313},
    and~Ching~Eng~Png\orcidlink{0000-0002-7797-1863}
\thanks{
This work was supported in part by the Agency for Science, Technology and Research (A*STAR) under Grant No.~C210917001.}%
\thanks{Wei-Bin~Ewe, Dax~Enshan~Koh, Siong~Thye~Goh, Hong-Son~Chu and Ching~Eng~Png are with the
Institute of High Performance Computing, Agency for Science, Technology and Research (A*STAR), 1 Fusionopolis Way, \#16-16 Connexis, Singapore 138632, Singapore (e-mail: ewewb@ihpc.a-star.edu.sg; dax\_koh@ihpc.a-star.edu.sg; gohst2@ihpc.a-star.edu.sg; chuhs@ihpc.a-star.edu.sg; pngce@ihpc.a-star.edu.sg).

\textcopyright\ 2022 IEEE. Personal use of this material is permitted. Permission from IEEE must be obtained for all other uses, in any current or future media, including reprinting/republishing this material for advertising or promotional purposes, creating new collective works, for resale or redistribution to servers or lists, or reuse of any copyrighted component of this work in other works.
}%
}

\maketitle

\begin{abstract}
Variational quantum algorithms are one of the most promising methods that can be implemented on noisy intermediate-scale quantum (NISQ) machines to achieve a quantum advantage over classical computers. This \papertype{} describes the use of a variational quantum algorithm in conjunction with the finite difference method for the calculation of propagation modes of an electromagnetic wave in a hollow metallic waveguide. The two-dimensional (2D) waveguide problem, described by the Helmholtz equation, is approximated by a system of linear equations, whose solutions are expressed in terms of simple quantum expectation values that can be evaluated efficiently on quantum hardware. Numerical examples are presented to validate the proposed method for solving 2D waveguide problems.
\end{abstract}


\section{Introduction}

\IEEEPARstart{Q}{uantum} computing is an emerging computing paradigm that seeks to take advantage of superposition and entanglement in quantum mechanics to improve computational efficiency and overcome the limitations of classical computing \cite{national2019quantum}. Over the last few decades, it has attracted the attention of many a researcher from a myriad of fields, including physics, chemistry, biology, computer science, and engineering. The potential acceleration promised by quantum algorithms is especially attractive to computer-aided engineering because of the growing need to solve more complex and larger-scale computational problems within a reasonable timeframe.

In 2009, Harrow, Hassidim, and Lloyd (HHL) proposed a quantum algorithm for solving systems of linear equations \cite{harrow2009quantum}. Their eponymous algorithm promises an exponential speedup over classical algorithms when the linear system is sparse and has a small condition number, and if the desired output is the result of some measurement of the solution vector instead of a classical description of the entire solution vector. Since then, a number of works have implemented \cite{cao2012quantum} and generalized \cite{clader2013preconditioned} the HHL algorithm and applied it to problems in various fields, for example, fluid dynamics \cite{steijl2018parallel,steijl2020quantum}, \added{electromagnetic problems\mbox{\cite{zhang2021quantum},}} semiconductors \cite{ morrell2021study}, and differential equations \cite{wang2020quantum}, including nonlinear differential equations \cite{Liu2021nonlinear,lloyd2020nonlinear}. A drawback, however, of the HHL algorithm is that the number of qubits and circuit depths needed to implement it for many problems of practical value are too large for current and near-term quantum computers. It may be many more years before fault-tolerant quantum computers capable of running the HHL algorithm become available.

In the meantime, recently developed cloud-based quantum hardware from IBM \cite{IBM}, Rigetti \cite{rigetti}, etc., have provided researchers with the opportunity to empirically implement quantum algorithms that use a limited number of qubits and have limited circuit depths. With this opportunity, a new class of quantum algorithms, called variational quantum algorithms (VQAs), has become popular and has emerged as one of the most promising candidates to achieve a practical quantum advantage over classical algorithms \cite{cerezo2021variational,endo2021hybrid,bharti2021towards,bharti2022noisy}. These VQAs are hybrid quantum-classical algorithms that take into account the limitations of current and near-term quantum hardware and distribute computational tasks between classical and quantum computers on the basis that some tasks can be run on one type of device more efficiently than on the other \added{(for a more comprehensive treatment of hybrid quantum-classical algorithms, we refer the reader to\mbox{\cite{cerezo2021variational}})}. In a VQA, the computational problem to be solved is encoded into a cost function, which is then expressed in terms of expectation values of Hamiltonians. An ansatz\footnote{\added{In the context of variational quantum circuits, an ansatz, also called a \textit{parameterized quantum circuit}, typically describes a subroutine consisting of a sequence of gates with tunable parameters applied to specific wires\mbox{\cite{cerezo2021variational}.}
}} with tunable parameters \deleted{(also called a \textit{parameterized quantum circuit})} is chosen and expectation values with respect to the ansatz are computed using a quantum computer. \replaced{The measurement values that are obtained from the quantum computer are fed into a classical optimizer, which updates the parameters of the ansatz}{The parameters of the ansatz are updated by a classical optimizer} in an outer loop that seeks to optimize the cost function iteratively. Popular VQAs include the variational quantum eigensolver (VQE), which computes the ground state energies of Hamiltonians \cite{peruzzo2014variational, mcclean2016theory} and the quantum approximate optimization algorithm (QAOA) \cite{farhi2014quantum}, which finds approximate solutions to combinatorial optimization problems. More recently, VQAs have been proposed to solve linear systems \cite{bravo2019variational, huang2019near,xu2021variational} and partial differential equations  \cite{lubasch2020variational,garcia2021solving,liu2021variational, sato2021variational,joo2021quantum}.

In this \papertype{}, we describe a novel approach to solve a microwave waveguide modes problem that can be implemented on current and near-term quantum computers. In particular, we design a variational quantum-based algorithm to calculate the propagation modes of an electromagnetic wave in a hollow metallic waveguide. Similar to HHL, the output of our algorithm is a quantum state, which can be measured to extract properties of the propagation modes. The wave propagation problem can be described by Helmholtz equations that can be approximated by eigenvalue equations via the finite difference method. Using a penalty method (called variational quantum deflation) introduced by \cite{higgott2019variational} and decomposition techniques from \cite{sato2021variational}, the eigenvalues and corresponding eigenvectors can be obtained by minimizing a cost function that is expressed in terms of quantum expectation values of simple Hamiltonians that can be evaluated efficiently on near-term quantum hardware. These eigenvectors are quantum states whose vector representation gives the solution to the waveguide modes problem.

The remainder of this \papertype{} is organized as follows. In Sec.~\ref{sec:formulation}, we describe a formulation of the problem of wave propagation in a metallic waveguide and use the finite difference method to approximate the problem by a system of linear equations. We then present an efficient decomposition of the linear system and express our cost function in terms of simple quantum expectation values. In Sec.~\ref{implementation}, we describe our algorithm for finding the propagation modes of the microwave waveguide and discuss details of our numerical implementation of it. In Sec.~\ref{results}, we present numerical results obtained by implementing our proposed algorithm on Qiskit's \textit{statevector simulator} \cite{qiskit}. Finally, we present a summary and discussion of our main results in Sec.~\ref{conclusion}.

\section{Formulation}
\label{sec:formulation}

\subsection{Microwave waveguide}

Consider a rectangular metallic hollow waveguide whose waveguide axis coincides with the $z$-axis and which is filled with a homogeneous material with permeability $\mu$ and permittivity $\varepsilon$. The electric field $E$ and magnetic field $H$ propagating in the waveguide can be decomposed into components along the $z$ axis and components transverse to the $z$ axis. Denoting the transverse electric and magnetic fields by $E_z$ and $H_z$ respectively, the two-dimensional scalar wave equations for the transverse electric (TE) wave and transverse magnetic (TM) wave can be obtained, respectively, as the following Helmholtz equations \cite[p.~37]{chew2021lectures}:
\begin{align}
\nabla_s^2 H_z &= - k_s^2 H_z,\quad \text{for\ TE\ waves}, 
\label{TEwave}
\\
\nabla_{s}^{2}E_{z} 
&= - k_{s}^{2}E_{z},\quad \text{for\ TM\ waves}
\label{TMwave} 
\end{align}
where the subscript $s$ represents the subspace transverse to the $z$-direction, and $k_s^2 = k^2 - k_z^2$, where $k=\sqrt{\omega^2\mu\varepsilon}$ is the wavenumber inside the waveguide---where $\omega$ is the angular frequency---and $k_z$ is the component of $k$ in the $z$-direction. The TE and TM mode waves satisfy the following boundary conditions on the metallic waveguide wall: the $E_{z}$ component satisfies the Dirichlet boundary condition $E_z = 0$, whereas the $H_z$ component satisfies the Neumann boundary condition $\hat{\mathrm{n}} \cdot \nabla_{s}H_{z} = \frac{\partial}{\partial \mathrm{n}}{H}_{z} = 0$, where $\hat{\mathrm{n}}$ denotes a unit vector normal to the boundary.

One could employ numerical methods to discretize \eqref{TEwave} and \eqref{TMwave} and convert them into matrix equations, which could then be solved on a classical computer. In this \papertype{}, however, we consider a different approach that allows the problem to be solved on near-term quantum computers. \replaced{Unlike the case in classical computation where $O(N)$ bits are required to represent a matrix equation of dimension $N$, here only $\log_2 N$ qubits are required. }{A potential benefit of this approach is an exponential reduction in the resources needed: a matrix equation of dimension $N$ can be represented by a quantum system of only $\log_2 N$ qubits.} First, the two-dimensional problem is cast as two separate, orthogonal one-dimensional wave problems, along the $x$-axis and along the $y$-axis. For a one-dimensional field distribution along the $x$-axis, the Laplacian operator in \eqref{TEwave} and \eqref{TMwave} simplifies to the second-order derivative $\frac{d^{2}}{dx^{2}}$, which can be approximated by using the central finite difference method as $\frac{d^{2}f}{dx^{2}} \approx \frac{ f_{x - 1} - 2f_x + f_{x + 1} }{\Delta x^{2}}$.
This approximation converts the scalar wave problem into an eigenvalue equation 
\begin{align}
   Mv = \lambda v,
   \label{eq:eigenvalue}
\end{align}
where $M$ is a matrix that depends on the boundary conditions, and $v$ and $\lambda$ denote its eigenvectors and eigenvalues respectively. More precisely, $M$ can be obtained by discretizing the cross section of the rectangular waveguide with a uniform shifted grid \cite{sarkar1989computation}: the matrix $M$ under Dirichlet boundary conditions (denoted $M_{x,D}$) and under Neumann boundary conditions (denoted $M_{x,N}$) can be obtained, respectively, as
\begin{align}
M_{x,D} = \frac{1}{\Delta x^{2}}
\begin{bmatrix}
3 & -1 & 0 & & & \cdots &  & 0 \\
-1 & 2 & -1 & 0 & & \cdots  & & 0 \\
0 & -1 & 2 & -1 & & \cdots  & & 0 \\
\vdots &  &  & \ddots  & & & & \vdots \\
0 & &  \cdots & & 0 & -1 & 2 & -1  \\
0 & &  \cdots & & & 0 & -1 & 3
\end{bmatrix}
\label{eq:MxD}
\end{align}

and
\begin{align}
M_{x,N} = \frac{1}{\Delta x^2}
\begin{bmatrix}
1 & -1 & 0 & & & \cdots &  & 0 \\
-1 & 2 & -1 & 0 & & \cdots  & & 0 \\
0 & -1 & 2 & -1 & & \cdots  & & 0 \\
\vdots &  &  & \ddots  & & & & \vdots \\
0 & & \cdots & & 0 & -1 & 2 & -1  \\
0 & & \cdots & & & 0 & -1 & 1
\end{bmatrix}.
\label{eq:MxN}
\end{align}
\added{For a derivation of \eqref{eq:MxD} and \eqref{eq:MxN}, we refer the reader to Appendix \ref{app:matrixM}.} The same procedure can be carried out for the one-dimensional discretization along the $y$-axis of a waveguide to obtain the matrices $M_{y,D}$ and $M_{y,N}$. We denote the number of grid points along the $x$ and $y$ directions by $2^{n_x}$ and $2^{n_y}$ respectively, where we have taken these numbers to be powers of 2 in order for the quantum algorithm that we will later describe to be implementable on multi-qubit systems. Note that we do not require that $n_x=n_y$, i.e.~the discretization sizes along $x$ and $y$ components could be different. For the simplicity of subsequent discussions, we shall assume that the grid size $\Delta x = 1$.

Next, the matrices $M_{x,j}$ and $M_{y,j}$ (for $j\in \{D,N\}$) can be combined using the Kronecker product $\otimes$ to produce matrices for the 2D waveguide modes\added{\mbox{\cite{Poisson2D}}}: the matrix for TM modes is given by
\begin{align}
  M_{\text{TM}} = I^{\otimes n_{y}} \otimes M_{x,D} + M_{y,D} \otimes I^{\otimes n_{x}} 
  \label{eq:MTM_decomp}
\end{align}
and the matrix for TE modes is given by
\begin{align}
  M_{\text{TE}} = I^{\otimes n_y} \otimes M_{x,N} + M_{y,N} \otimes I^{\otimes n_x}, 
 \label{eq:MTE_decomp}
\end{align}
where $I$ denotes the $2\times 2$ identity matrix\added{, and the Kronecker product symbol on the exponent denotes iterated Kronecker multiplication: for example, $A^{\otimes k} = \underbrace{A \otimes A \otimes \ldots\otimes A}_{\scriptsize \mbox{$k$ times}}$.}

In this \papertype{}, we restrict our attention to the case where $M \in \{M_{\text{TE}}, M_{\text{TM}}\}$. The propagation modes of the microwave waveguide are encoded by the solution vectors $v$ of the eigenvalue equation \eqref{eq:eigenvalue} with the above $M$'s. To distinguish between the different propagation modes, we denote the eigenvalues of $M$ by $E_i$ and their corresponding normalized eigenvectors by the state vectors $\ket{v_i}$, where we have used Dirac's bra-ket notation \cite{dirac1939new}. We use the convention of ordering the eigenvalues in increasing order: $E_0 \leq E_1 \leq \ldots \leq E_{2^n-1}$, where $n=n_x+n_y$.

\subsection{Cost function}\label{cost-function}

Our goal is to solve the eigenvalue equation $M\ket {v_i} = E_i \ket {v_i}$,  for $M \in \{M_{\text{TE}}, M_{\text{TM}}\}$. We start by describing the task of finding the ground state energy $E_0$ of $M$. Consider the cost function \begin{align}
    F_0(\theta) = \bra {\psi(\theta)}M \ket{\psi(\theta)},
    \label{eq:groundStateVariationalEnergy}
\end{align}
where $\ket{\psi(\theta)} \in \mathcal A$ is an ansatz chosen from a family $\mathcal A$ of ansatzes parameterized by $\theta$. For any $\theta$, the cost function \eqref{eq:groundStateVariationalEnergy} gives an upper bound for the ground state energy, i.e.~$F_0(\theta) \geq E_0$. Furthermore, if $\mathcal A$ is a sufficiently rich class, then minimizing \eqref{eq:groundStateVariationalEnergy} with respect to $\theta$ by using, say, a variational quantum algorithm will yield a good approximation of $E_0$ \cite{peruzzo2014variational}.

Besides computing the lowest propagation mode of the microwave waveguide, the variational approach can also be used to compute the second- and higher-order eigenvalues to determine the cut-off frequencies and ensure operation in a single-mode environment. In order to compute the $k$-th excited-state energy $E_k$ of the matrix $M$, the cost function \eqref{eq:groundStateVariationalEnergy} can be modified by including additional penalty terms:
\begin{align}
  F_k(\theta) = \bra{\psi(\theta)}  M \ket{\psi(\theta)} + \sum_{i = 0}^{k - 1}{\beta_{i}\left|
  \bket{\psi(\theta)}{\psi(\theta^{(i)})}
  \right|^{2}}, 
  \label{eq:higher_order} 
\end{align}
where for each $i\in\{0,\ldots,k-1\}$, \(\beta_{i}\) is chosen to be any large constant satisfying $\beta_{i} > E_k - E_i$, and $\theta^{(i)}$ is defined as the (previously-found) angle that minimizes (or approximately minimizes) the cost function $F_i$, i.e.
$F_i(\theta^{(i)}) \approx \min_\theta F_i(\theta)$ \cite{higgott2019variational}. The minimizations of \eqref{eq:higher_order}  are performed iteratively, starting from $k=0$ and increasing $k$ by one in each iteration until all the desired higher-order eigenvalues are obtained. Note that the second term in \eqref{eq:higher_order} enforces the constraint that the state $\ket{\psi( \theta^{(k)})}$ minimizing the cost function is orthogonal to the previously optimized states
$\ket{\psi(\theta^{(0)})}, \ldots, \ket{\psi(\theta^{(k-1)})}$.

In order to evaluate the expectation value in \eqref{eq:higher_order} using quantum circuits (over typical gate sets), we need to provide a decomposition of $M$ into simpler observables. A canonical way to achieve this is to decompose $M$ over the Pauli basis $\mathcal P_n = \big\{ P_1 \otimes \ldots \otimes P_n : \forall i,\ P_i \in \{I, X, Y, Z\}\big\}$ as
\begin{align}
   M = \sum_{P \in \mathcal P_n}{c_P P} ,
   \label{eq:pauli_expansion_M}
\end{align}
where $c_P = \frac{1}{2^n}\tr(PM)$ are the scalar coefficients in the expansion and
\begin{align}
X=\kbra 10+\kbra 01,\ Z=\kbra 00-\kbra 11,\ Y=iXZ   
\label{eq:PauliMatrices}
\end{align}
are the Pauli matrices \cite{nielsen2010quantum}. A major drawback of this decomposition though is that the number of non-zero coefficients $c_P$ in \eqref{eq:pauli_expansion_M}---and hence the number of expectation values that would need to be evaluated---increases exponentially with $n$.

To circumvent this drawback, we make use of a more efficient decomposition similar to that proposed recently by Sato \emph{et al.} \cite{sato2021variational} that expresses $M$ as a linear combination of unitary transformations of simple Hamiltonians. This decomposition has the advantage that the number of terms in the decomposition---and hence the number of expectation values that would need to be evaluated---is a constant independent of $n$. To decompose $M$, note that we could write the matrices $M_{x,D}$ from \eqref{eq:MxD} and $M_{x,N}$ from \eqref{eq:MxN}, as well as the matrices $M_{y,D}$ and $M_{y,N}$, as
\begin{align} 
M_{t,j} &= P_{n_t}^\dag \left[ I^{\otimes n_t - 1}\otimes\left( I - X \right) + I_{0}^{\otimes n_t - 1}\otimes\left( X + a_j I \right) \right] P_{n_t} \nonumber\\ &\quad + I^{\otimes n_t - 1}\otimes\left( I - X \right),
\label{eq:rewritecombine}
\end{align}
for $t \in \{x,y\}$ and $j \in \{D,N\}$, where $a_D = 1$ and $a_N = -1$; $I_0 = \kbra 00$; and $P_n$ denotes the $n$-qubit cyclic shift operator defined by
\begin{align}
P_n = \sum_{i=0}^{2^n-1}
\kbra{(i+1)\ \mathrm{mod}\ 2^n}i
.
\label{eq:shiftP}
\end{align}

By substituting \eqref{eq:rewritecombine} into \eqref{eq:MTM_decomp} and \eqref{eq:MTE_decomp} and simplifying the resulting expressions, we find that for $j \in \{\text{TM},\text{TE}\}$,
\begin{align}
    M_j = 4 I^{\otimes n_y+n_x}
    +
    \sum_{i=1}^2 H_i
    +
    \sum_{i=3}^5 V^\dag H_i V
    +
    \sum_{i=6}^8 W^\dag H_i W,
    \label{eq:longcombinedM}
\end{align}
where
\begin{align}
   V &= I^{\otimes n_y} \otimes P_{n_x},
   \nonumber\\
    W &= P_{n_y} \otimes I^{\otimes n_x},
    \nonumber\\
    H_1 &= H_3 = - I^{\otimes n_y+n_x-1} \otimes X,
    \nonumber\\
    H_2 &= H_6 = - I^{\otimes n_y-1} \otimes X \otimes I^{\otimes n_x}, 
    \nonumber\\
    H_4 &= I^{\otimes n_y} \otimes I_0^{\otimes n_x-1} \otimes X,
    \nonumber\\
    H_5 &= b_j I^{\otimes n_y} \otimes I_0^{\otimes n_x-1} \otimes I,
    \nonumber\\
    H_7 &= I_0^{\otimes n_y-1} \otimes X \otimes I^{\otimes n_x},
    \nonumber\\
    H_8 &= b_j I_0^{\otimes n_y-1} \otimes I^{\otimes n_x+1},
\end{align}
and $b_{\text{TM}} = 1$ and $b_{\text{TE}}  = -1$. The decomposition \eqref{eq:longcombinedM} expresses $M_j$ as a sum of unitary conjugations of simple Hamiltonians $4 I^{\otimes n_y+n_x}, H_1, \ldots, H_8$ and allows the cost function \eqref{eq:higher_order} to be written in terms of expectation values that can be evaluated by simple measurements of quantum states prepared on a quantum computer. In particular, the first term of \eqref{eq:higher_order}, with $M=M_j$, for $j \in \{\text{TM},\text{TE}\}$, can be written as
\begin{align}
    \bra{\psi(\theta)}  M_j \ket{\psi(\theta)}
    &=
    4 + \sum_{i=1}^8 
    \bra{\phi_i(\theta)}  H_i \ket{\phi_i(\theta)},
    \label{eq:efficientDecomposition}
\end{align}
where $\ket{\phi_i(\theta)} = \ket{\psi(\theta)}$ for $i=1,2$; $\ket{\phi_i(\theta)} = V\ket{\psi(\theta)}$ for $i=3,4,5$; and
$\ket{\phi_i(\theta)} = W \ket{\psi(\theta)}$ for $i=6,7,8$. Hence, unlike the Pauli decomposition \eqref{eq:pauli_expansion_M} where the number of expectation values grows exponentially in $n$, the decomposition presented in \eqref{eq:efficientDecomposition} involves only a constant (in $n$) number of expectation values.

\section{Implementation}\label{implementation}

\begin{figure}
      \centering
  \begin{tikzpicture}[node distance=1.8cm,auto]
  \node (start) [startstop] {Start};
  \node (in1) [io, below of=start] {Initalize $\theta^{(k)}= \theta_0^{(k)}$};
  \node (pro1) [process, below of=in1] {Evaluate $F_k(\theta^{(k)})$};
\node (dec1) [decision, below of=pro1, yshift=-0.6cm] {Terminate?};
\node (pro2b) [process, right of=dec1, xshift=2cm, yshift= 1.2cm] {Update $\theta^{(k)}=\theta^{(k)}_{\text{new}}$};
\node (out1) [io, below of=dec1] {Output: $\theta^{(k)}$};

\draw [arrow] (start) -- (in1);
\draw [arrow] (in1) -- (pro1);
\draw [arrow] (pro1) -- (dec1);
\draw [arrow] (dec1) -- node {yes}(out1);
\draw [arrow, anchor=left, above] (dec1) -| node [xshift=-2cm]{no} (pro2b);
\draw [arrow] (pro2b) |- (pro1);
  \end{tikzpicture}
\caption{Flowchart describing our algorithm. We carry out the above procedure for $k=0, \ldots, m-1$ in sequence.}
\label{fig:flowchartAlgo}
\end{figure}
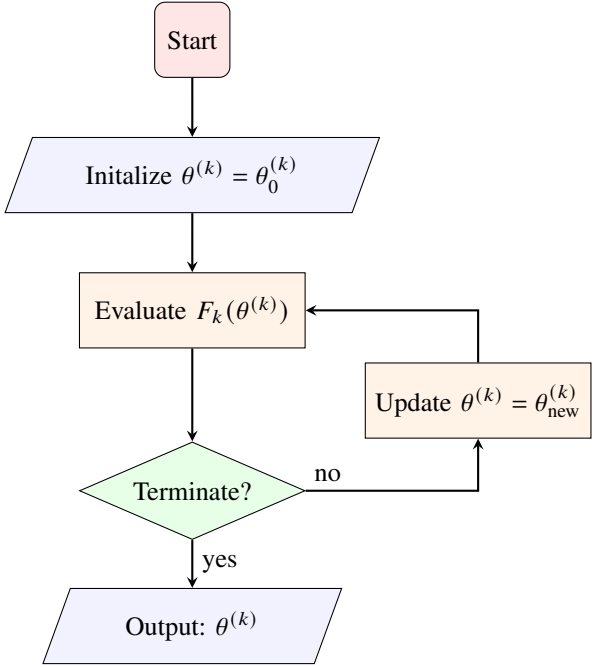

In this section, we shall describe our proposed algorithm for solving the 2D waveguide modes
problem. To obtain the first $m$ modes of the microwave waveguide, we perform the following algorithm (see Fig.~\ref{fig:flowchartAlgo}):
\begin{enumerate}
\def\labelenumi{\arabic{enumi})}
\item
  Set $k=0$.
\item 
  Initialize a set of parameters $\theta$ on a classical computer.
\item
  Evaluate the cost function $F_k(\theta)$ in \eqref{eq:higher_order} on a quantum computer.
\item
  Set $\theta^{(k)} = \theta$ and proceed to the next step if one of the terminal conditions is
  satisfied; otherwise, update the set of parameters \(\theta\) using some classical optimization scheme and return to step 3.
\item
  If $k=m-1$, halt. Otherwise, increment $k$ by 1 and return to step 2 for the next waveguide mode.
\end{enumerate}

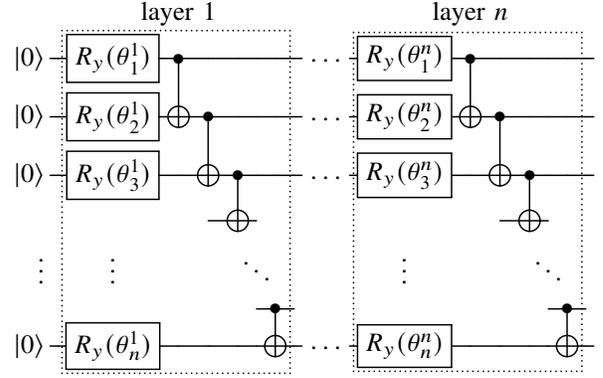
\begin{figure}
\begin{align*}
\Qcircuit @C=0.3em @R=.4em {
 &  & & \mbox{layer 1} 
& &  & & & & & & & &  &  & & & \mbox{layer $n$}  
\\~\\
& \lstick{\ket{0}}\qw & \gate{R_y(\theta^1_1)} & \ctrl{1} & \qw      & \qw   & \qw & \qw & \qw & \qw &  & &  \ldots & & &   &  \gate{R_y(\theta^n_1)} & \ctrl{1} & \qw      & \qw   & \qw & \qw & \qw & \qw
\\
& \lstick{\ket{0}}\qw & \gate{R_y(\theta^1_2)} & \targ    & \ctrl{1} & \qw   & \qw & \qw & \qw & \qw &  & &  \ldots & &  & & \gate{R_y(\theta^n_2)} & \targ    & \ctrl{1} & \qw   & \qw & \qw & \qw & \qw
\\
& \lstick{\ket{0}}\qw & \gate{R_y(\theta^1_3)} & \qw & \targ    &   \ctrl{1} & \qw & \qw & \qw & \qw &  & & \ldots & & & &   \gate{R_y(\theta^n_3)} & \qw      & \targ    &   \ctrl{1} & \qw & \qw & \qw & \qw 
\\
& &  & & & \targ & \qw & & & & & & & & &  & & & & \targ & \qw & 
\\~\\~\\
& \lstick{\vdots} & \vdots & & &  &  \ddots  & & & & & & & & & & \vdots  & & & & \ddots &
\\~\\~\\~\\
& & & & & & &  \ctrl{1} & \qw  & & & & & & & &  & & & & & \ctrl{1} & \qw &
\\
& \lstick{\ket{0}}\qw & \gate{R_y(\theta^1_n)} & \qw      & \qw    &  \qw & \qw & \targ & \qw & \qw &  &  & \ldots & & & &   \gate{R_y(\theta^n_n)} & \qw      & \qw    &  \qw & \qw & \targ & \qw & \qw 
\gategroup{3}{3}{14}{8}{.3em}{.}
\gategroup{3}{17}{14}{22}{.3em}{.}\\~\\
}
\end{align*}\caption{Circuit diagram of the hardware-efficient ansatz used in this \papertype{}. The ansatz comprises $n$ layers acting on an $n$-qubit register initialized to the computational basis state $\ket 0^{\otimes n}$, where each layer 
consists of a single-qubit rotation gate $R_y(\theta_i) = \exp(-i \theta_i Y/2)$ acting on each register followed by controlled-NOT gates arranged according to a linear entanglement structure.
}
\label{fig:ansatz}
\end{figure}

The trial quantum state
\(\left| \psi(\theta) \right\rangle\) in step 2 is prepared
by applying a parametrized unitary $U(\theta)$ to the initial state
$\ket {0^{\otimes n}}$. In this \papertype{}, the state $U(\theta)\ket {0^{\otimes n}}$ is chosen to be the
hardware-efficient ansatz (HEA) shown in Fig.~\ref{fig:ansatz}, which consists of both rotation gates $R_y(\theta) = \exp(-i \theta Y/2)$ and controlled NOT gates, where the latter are arranged in a linear entanglement structure. To ensure that our ansatz is sufficiently expressive, we use a multi-layered HEA with $n=n_x+n_y$ layers.

To implement the cyclic shift operators \eqref{eq:shiftP} that are applied to the ansatz $\ket{\psi(\theta)}$ to obtain
$V\ket{\psi(\theta)}$ and $W\ket{\psi(\theta)}$ in \eqref{eq:efficientDecomposition}, we note that $P_k$ can be written as a product of $k$ multiple-control Toffoli gates, where the largest of these gates involves $k-1$ controls. More precisely, $P_k = \prod_{i=0}^{k-1} C^{(i)} (X)_{k,k-1,\ldots,k-i}$, where $C^{(i)} (X)_{k,k-1,\ldots,k-i}$ is the multiple-control Toffoli gate with $(k,k-1,\ldots,k-i+1)$ as the control qubits and $k-i$ as the target qubit (see, for example, \cite[Fig.~4]{zhang2014quantum}).
The multiple-control Toffoli gate can in turn be decomposed into a circuit consisting of only a linear number of $T$ gates, controlled-NOT gates and Hadamard gates, with a linear number of ancilla qubits that are each set to and returned to the computational basis state $\ket 0$ \cite{maslov2016advantages}. In other words, each $C^{(i)} (X)$ gate in the decomposition of $P_k$ increases the circuit depth (with respect to the Clifford+$T$ gate set) by $O(k)$ and uses $O(k)$ ancilla qubits that can be reused. This results in an overall circuit with $O(k)$ ancilla qubits and depth $O(k^2)$ that is needed to implement $P_k$. Hence, applying the cyclic shift operators using the above decomposition to the hardware-efficient ansatz gives an efficient way to prepare the states $\ket{\phi_i(\theta)}$ that can be measured on a quantum computer to evaluate the cost function \eqref{eq:higher_order}.

Step 4 of our algorithm involves a classical optimization subroutine that is used to minimize the cost function \eqref{eq:higher_order}. This subroutine could be performed by either gradient-based or gradient-free optimizers, which differ based on whether they make use of information about the gradient of the cost function. Unlike gradient-free optimizers, gradient-based optimizers utilize this information to find a good search direction that informs how the parameters in the variational circuit are updated. An example of a gradient-based optimizer is the Broyden-Fletcher-Goldfarb-Shanno (BFGS) optimizer \cite{broyden1970convergence, fletcher1970new,
goldfarb1970family, shanno1970conditioning}, which we will use in our simulations.

We conclude this section by giving an analytical expression for the gradient of the cost function that can be used by the gradient-based optimizer. By denoting quantum expectation values as $\big\langle H \big\rangle_{\phi} = \bra \phi H \ket \phi$, we find that the $j$th component of the gradient is given by
\begin{align}
    \frac{\partial F_k(\theta)}{\partial \theta_j}  
    &=
    \big\langle X\otimes M \big\rangle_{
    \partial_j \psi(\theta),\psi(\theta)
    } + 
    \sum_{i=0}^{k-1} \beta_i
    \big\langle X\otimes \kbra{0}{0} \big\rangle_{
    \Phi_{ij}(\theta)
    },
    \label{eq:dFdt}
\end{align}
where 
\begin{align}
    \ket{\partial_j \psi(\theta),\psi(\theta)} &=\frac{1}{\sqrt{2}}\left( \ket{0}\otimes\ket{\partial_j \psi(\theta)}+\ket{1}\otimes\ket{\psi(\theta)} \right), \nonumber\\
    \ket{\Phi_{ij}(\theta)} &= \left[ I \otimes U^\dag(\theta^{(i)})\right] \ket{\partial_j \psi(\theta),\psi(\theta)}\mbox{, and}
    \nonumber\\
    \ket{\partial_j \psi(\theta)} &= 
    2
    \frac{\partial \ket{ \psi(\theta)}}{\partial \theta_j}.
\end{align}
A detailed derivation of \eqref{eq:dFdt} is provided in \replaced{Appendix \ref{app:derivative}}{the Appendix}. Since \eqref{eq:dFdt} is expressed as a linear combination of quantum expectation values, the gradient of the cost function can be evaluated on a quantum computer. A potential disadvantage of such an approach, though, is that the number of shots needed to evaluate the gradient on a quantum computer might be large. In such a scenario, it might be preferable to approximate the gradient by means of numerical differentiation.

\section{Results}\label{results}

For our implementation, we consider the propagation modes of a 15 mm $\times$ 10 mm rectangular metallic waveguide. For the computation of the characteristics of the waveguide propagation, we implemented our proposed algorithm and ran the simulation using the \emph{statevector simulator} on Qiskit \cite{qiskit}, IBM's open-source framework for working with quantum circuits and algorithms \added{and coordinating between classical and quantum hardware}.
For the minimization process, the BFGS optimizer (see discussion in Sec.~\ref{implementation}) was deployed to update the ansatz parameter \(\theta\) in the iteration step.

\begin{table}[htp]
\centering
\caption{Comparison of VQA solution for TE and TM modes with classical and analytical solutions}
\label{table1}

\begin{tabular}{cccc}
\multirow{2}{*}{Mode} & VQA    & Classical solution & Analytical solution\\ 
       &  $f_{\text{cut-off}}$ (GHz)     & $f_{\text{cut-off}}$ (GHz)  & $f_{\text{cut-off}}$ (GHz) \\ 
\hline
$\text{TE}_{10}$ & 9.9770  & 9.9770     & 9.9931      \\
$\text{TE}_{01}$ & 14.8935 & 14.8935    & 14.9896     \\
$\text{TM}_{11}$ & 17.9265 & 17.9264    & 18.0153     \\
$\text{TM}_{21}$ & 24.8225 & 24.8225    & 24.9827     \\ 
\hline
\end{tabular}
\end{table}

\begin{figure}
      \centering
      \subcaptionbox{}
        {\includegraphics[width = 0.45\linewidth]{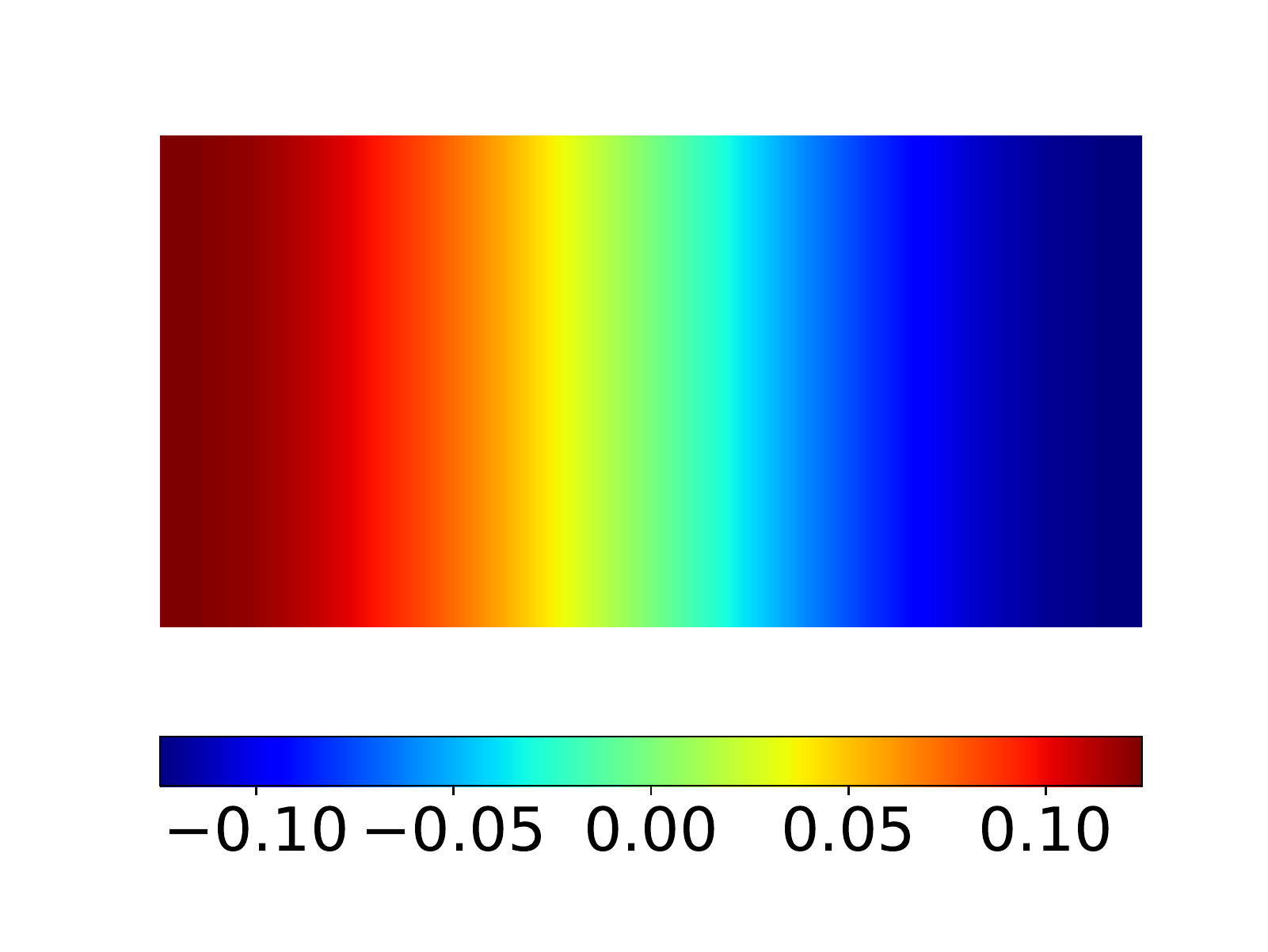}}
     \subcaptionbox{}
        {\includegraphics[width = 0.45\linewidth]{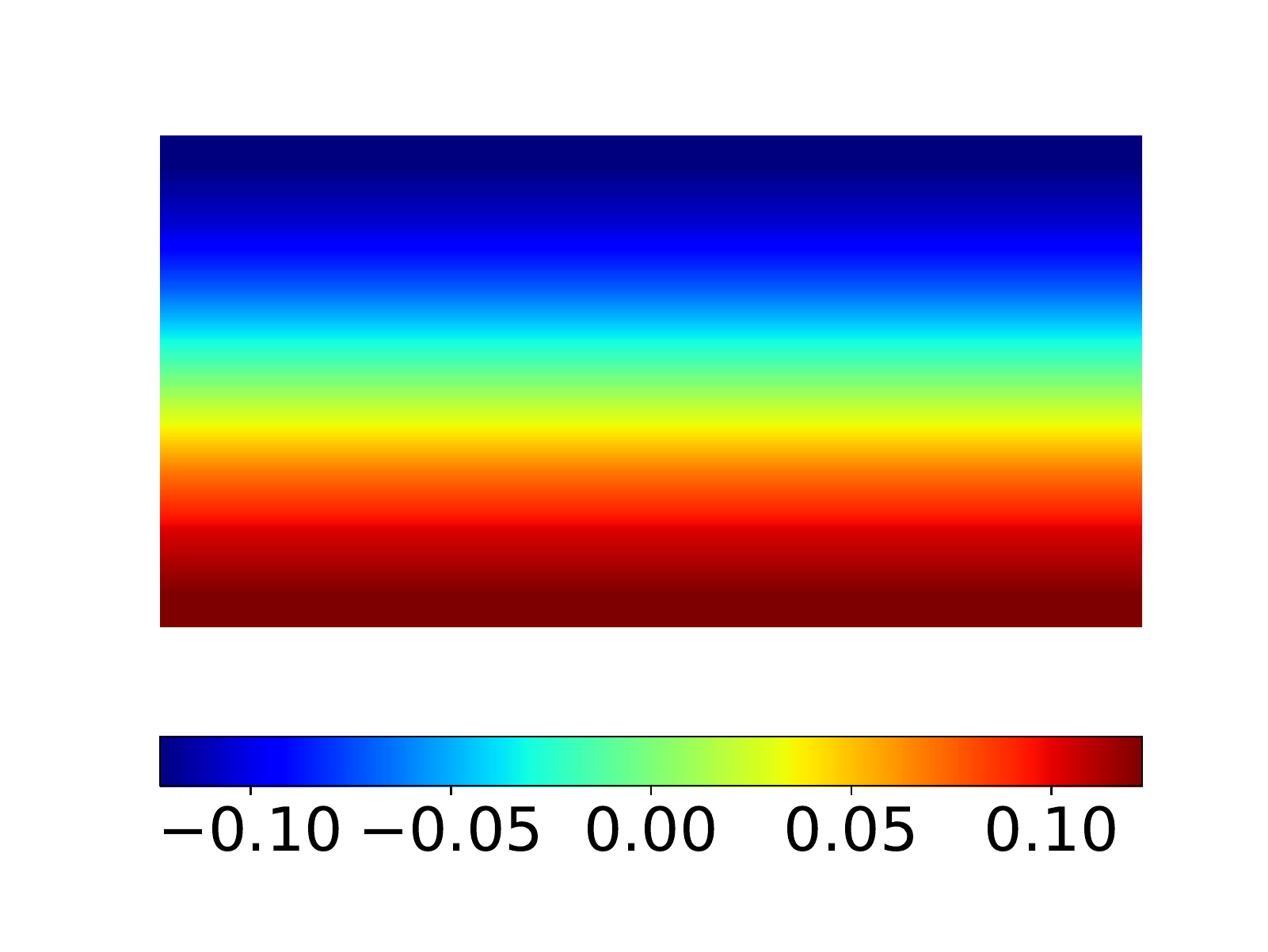}}
      \caption{Two-dimensional color plot of the $H_z$ of computed waveguide TE modes obtained from the VQA: (a) $\text{TE}_{10}$, and (b) $\text{TE}_{01}$.  }\label{fig:TEplot}
\end{figure}

\begin{figure}[htbp]
      \centering
      \subcaptionbox{}
        {\includegraphics[width = 0.45\linewidth]{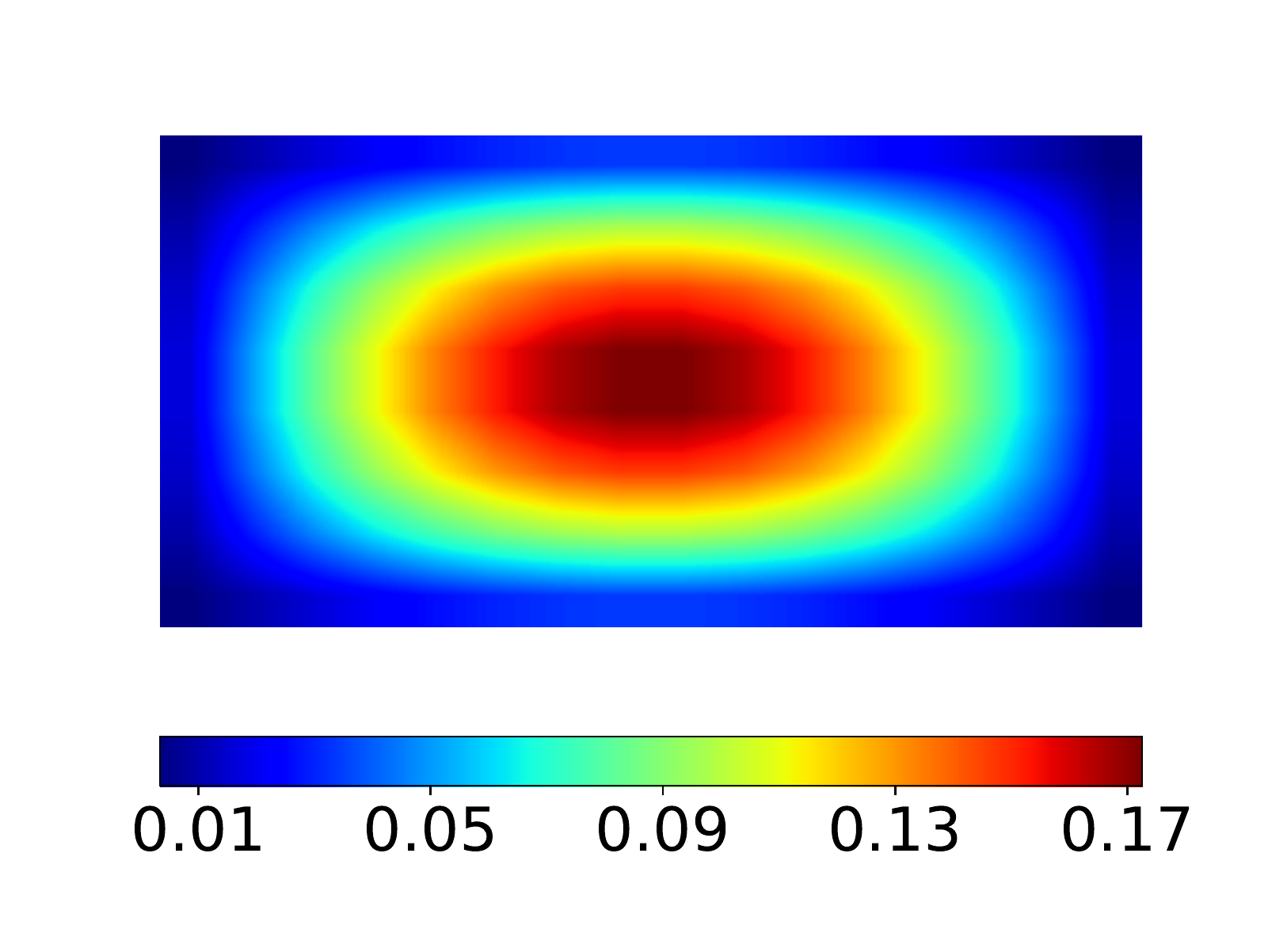}}
     \subcaptionbox{}
        {\includegraphics[width = 0.45\linewidth]{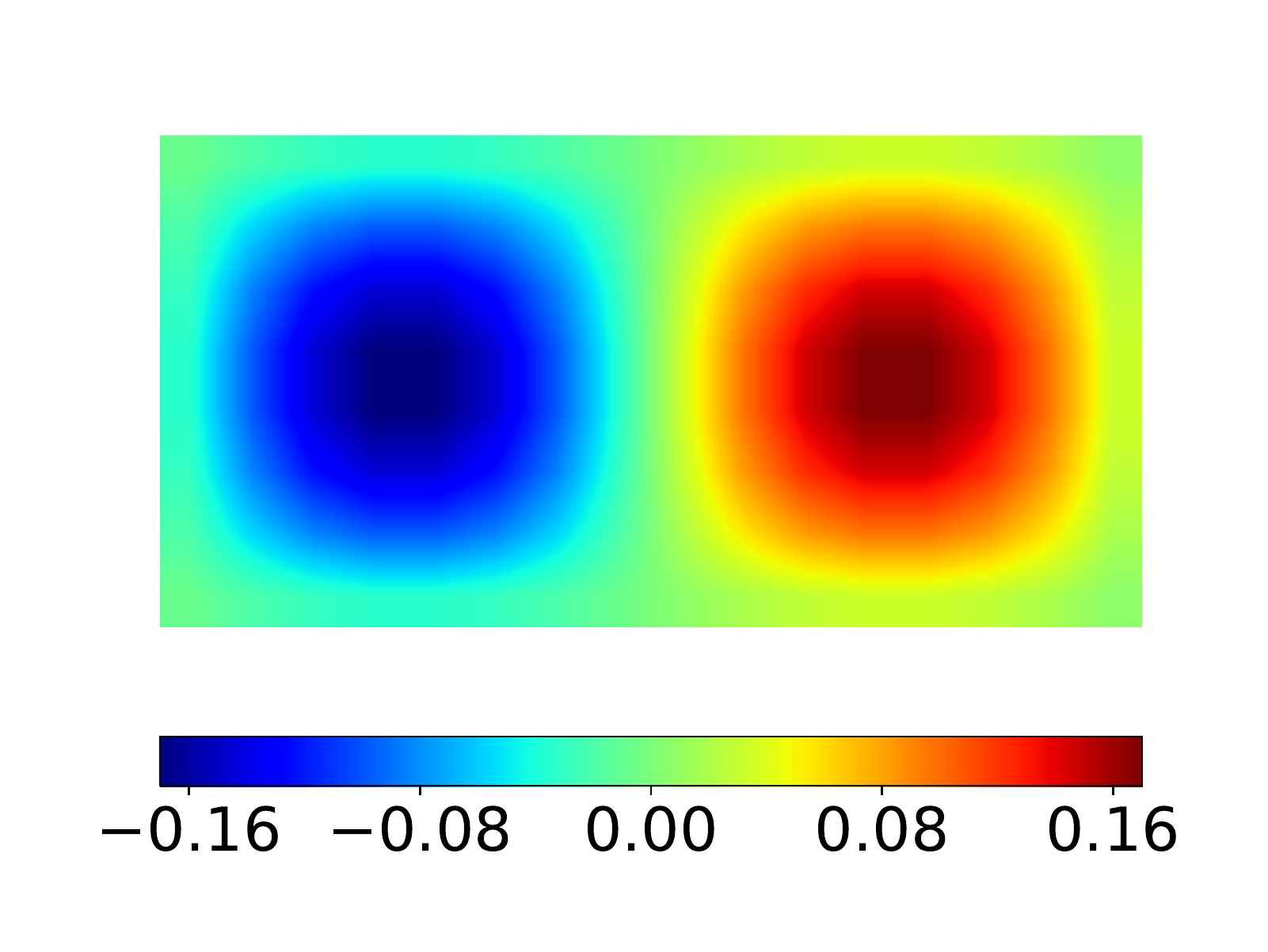}}
      \caption{Two-dimensional color plot of the $E_z$ of computed waveguide TM modes obtained from the VQA: (a) $\text{TM}_{11}$, and (b) $\text{TM}_{21}$.}\label{fig:TMplot}
\end{figure}

In Table \ref{table1}, the cut-off frequency of the first two TM and TE modes obtained from minimizing \eqref{eq:higher_order}  using VQE are tabulated and compared with the results of classical and analytical solutions. Here, the analytical solution refers to the exact solution obtained by solving the Helmholtz equations \eqref{TEwave} and \eqref{TMwave} directly, and the classical solution refers to the solution obtained by classically solving the eigenvalue equation \eqref{eq:eigenvalue} \added{using LAPACK routines}. The results for VQA are obtained by averaging the outcomes of five trials for  $n_x = 4$ and $n_y = 3$ qubits in the horizontal and vertical directions, respectively. For both TE and TM modes, the cut-off frequencies computed from the VQA are almost identical to the classical solutions with an error of below 0.001\%. When compared with the analytical solutions, the errors are below 1\%, which validates the accuracy of our proposed method. By using the optimized parameters computed in Table \ref{table1}, the field distributions of the corresponding modes can be reconstructed from the ansatz: the $H_{z}$ field distributions of the computed TE modes and the $E_{z}$ field distributions of the computed TM modes are shown in Figs.~\ref{fig:TEplot} and \ref{fig:TMplot} respectively.

\begin{figure}
\centering
  \includegraphics[width=0.9\linewidth]
  {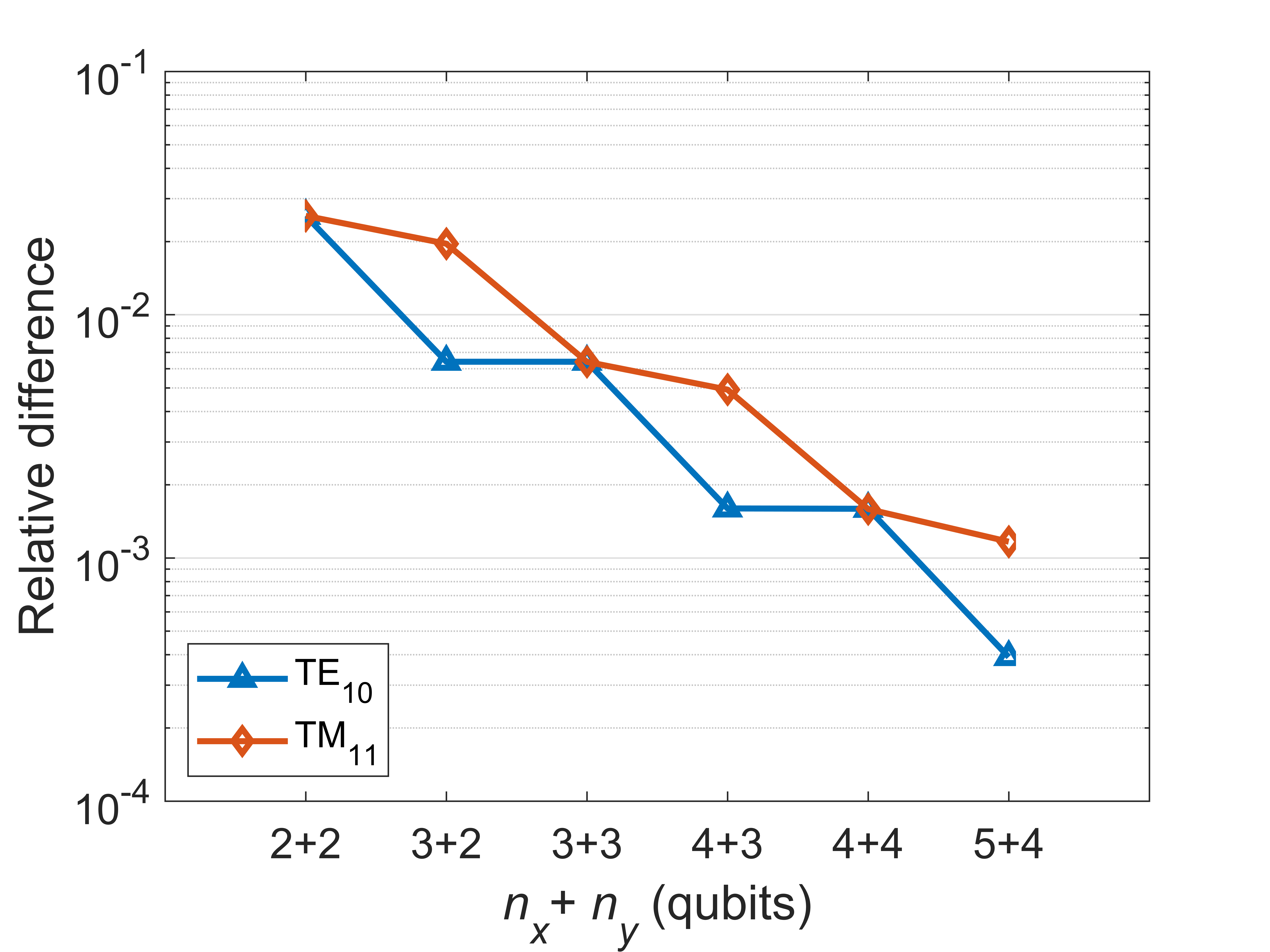}
\caption{The relative difference of cut-off frequency between the VQE results and analytical solution with an increasing number of $n_x$ and $n_y$ qubits.}
\label{fig:conv}
\end{figure}

Fig.~\ref{fig:conv} plots the relative difference between the VQA results and the analytical solution of the cut-off frequency  with an increasing number $n_x$ of qubits along the horizontal axis and an increasing number $n_y$ of qubits along the vertical axis, of the waveguide. In general, the results for both the TE and TM modes exhibit a logarithmic convergence rate with an increasing number of qubits deployed. It is found that the relative difference of the TE\textsubscript{10} results reduces by almost two orders of magnitude when \(n_{x}\) increases from two to five qubits while no improvement is observed when \(n_{y}\) increases. This is because the \(H_{z}\) field of TE\textsubscript{10} varies only horizontally; therefore, increasing the number \(n_{x}\) of qubits improves the sampling and captures the variation of $H_{z}$ more accurately. On the other hand, for the TM\textsubscript{11} mode, the $E_z$ field varies both horizontally and vertically. It is observed that the relative difference of the cut-off frequency for the TM\textsubscript{11} mode reduces when either $n_x$ or $n_y$ is increased.

\begin{figure}
\centering
      \subcaptionbox{}
        {\includegraphics[width = 0.9\linewidth]{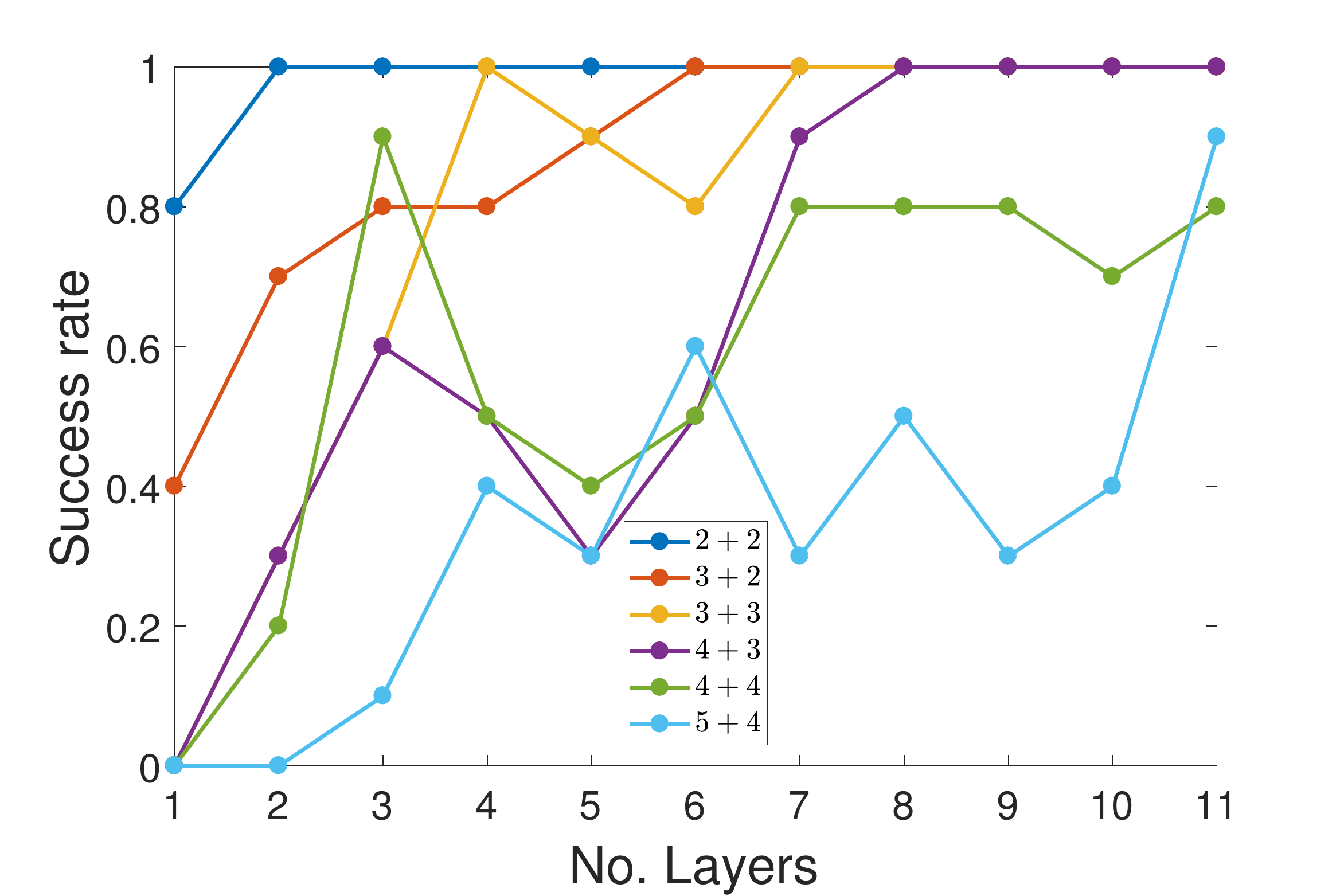}}
     \subcaptionbox{}
        {\includegraphics[width = 0.9\linewidth]{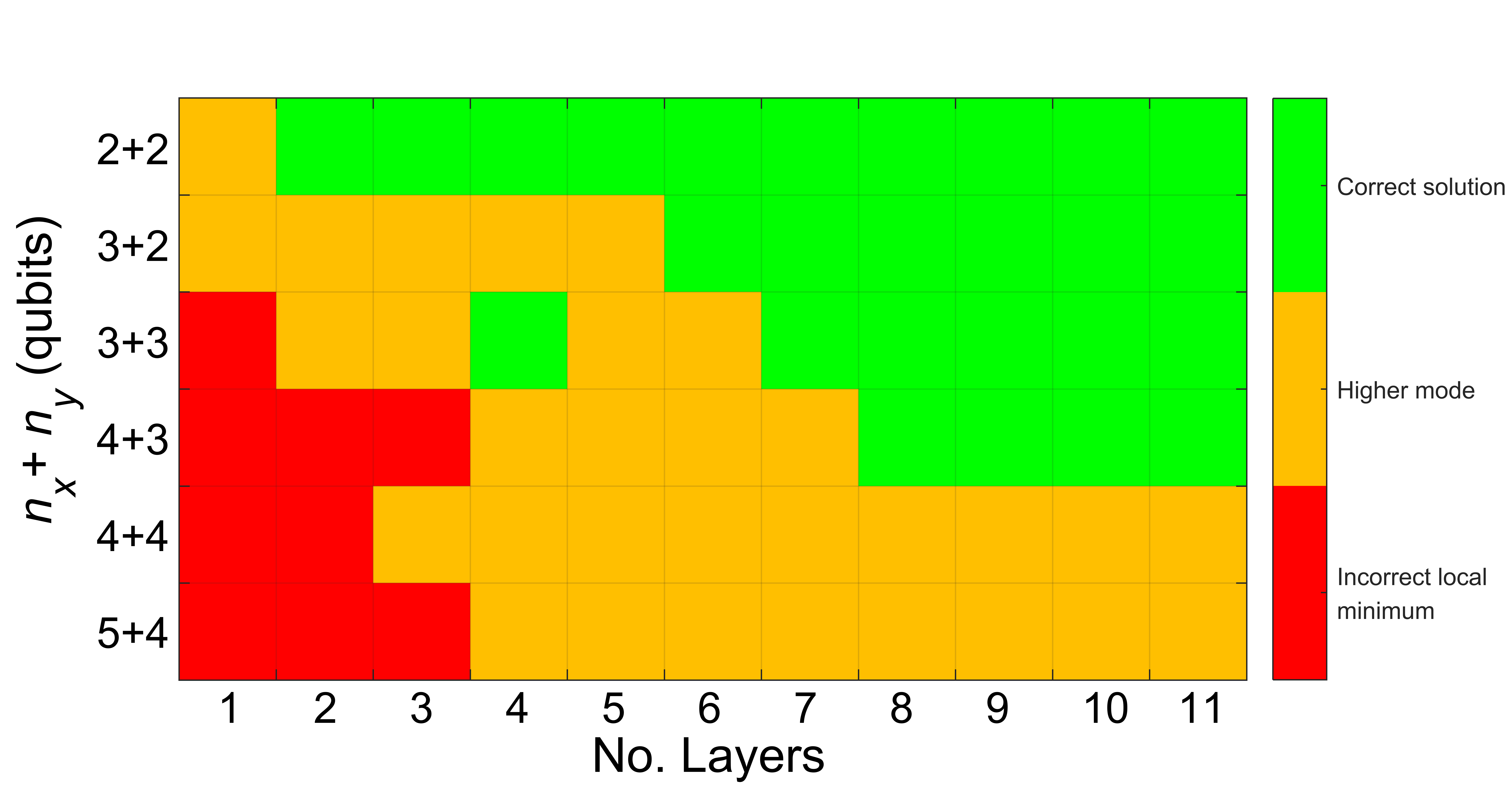}}
      \caption{ 
      The effect of different number of ansatz layers on finding the solution of TM$_{11}$ mode: (a) The success rates of finding correct solutions; and (b) Three types of solutions obtained -- correct solutions \replaced{({\color{green}green color})}{(green color)}, solutions that converged to higher modes \replaced{({\color{amber}amber color})}{(amber color)}, and solutions that converged to incorrect local minima \replaced{({\color{red}red color})}{(red color)}.}
      \label{fig:suc_rate}
\end{figure}

Lastly, the effects of changing the number of layers in the ansatz on finding the TM$_{11}$ mode solution are shown in Fig.~\ref{fig:suc_rate}. The results are obtained from the minimization of six different $n_x, n_y$ scenarios and up to 11 layers of HEA are considered due to resource constraints. The success rates shown in Fig.~\ref{fig:suc_rate}a measure the correct solutions that converged with fidelity $|\bket{x}{\psi(\theta)}|^2 \ge 0.95$, where $x$ is the eigenvector obtained from the classical algorithm. Fig.~\ref{fig:suc_rate}b tabulates the three different types of solutions obtained by using different multi-layered HEAs: \textit{green} means that the correct solutions were obtained in all trials; \textit{amber} means that a majority of the trials have solutions that converged to higher modes; and \textit{red} means that the majority of the trials have solutions that converged to incorrect minima. In general, the chances of getting the correct solution or the global minimum increases with the number of ansatz layers used because the expressibility of the ansatz---i.e.~the ability to explore the space of states---improves with increasing the number of parameters. 

\section{Conclusion}
\label{conclusion}

This \papertype{} has extended the use of variational quantum-based algorithms to the task of solving for the propagation modes of an electromagnetic wave in metallic waveguides. By using the proposed algorithm, the eigenvalue equation describing the 2D waveguide modes problem can be solved by performing quantum measurements of simple observables with respect to efficiently-preparable quantum states. We have shown that the number of expectations that need to be evaluated is independent of the system size. Numerical experiments have been presented to validate the proposed method for computing the propagation modes of a metallic waveguide. They demonstrate, inter alia, a significant improvement of accuracy when the number of qubits is only slightly increased.

For our implementation of the algorithm in this \papertype{}, we used the hardware-efficient ansatz with $R_y$ rotation gates and controlled-NOT entanglers. We leave open the question about the performance (e.g.~rate of convergence, success rate) of other ansatzes. Investigating, benchmarking, and proposing metrics to assess the performance of different ansatzes for various problems are currently active areas of research \cite{cerezo2021variational,Sim_2019, choquette2021quantum, you2021exploring}.

Our study has assumed that
the state preparations, unitary transformations and measurements in the quantum circuits used are implemented perfectly. However, real-world quantum computers are susceptible to noise, which would degrade the quality of the solutions obtained. An important question we leave open for future work is the extent to which noise affects the performance of our algorithm, and the degree to which error mitigation techniques \cite{li2017efficient,temme2017error,endo2018practical,endo2021hybrid,mari2021extending,takagi2021optimal} can reduce the effects of the noise. Such an endeavor should take into consideration some recent theoretical and numerical work that have highlighted some limitations of quantum error mitigation on expectation estimation and training quantum circuits \cite{takagi2021fundamental, wang2021error}.

\appendices
\section{1D Dirichlet and Neumann Matrices}
\label{app:matrixM}

\begin{figure}
\centering
  \includegraphics[width=0.5\linewidth]
  {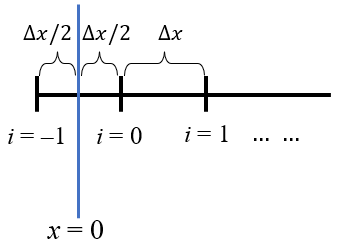}
\caption{\added{At the boundary of the 1D finite difference mesh based on uniform shifted grid\mbox{\cite{sarkar1989computation}}}
}
\label{fig:1DMesh}
\end{figure}

\addedStart

In this appendix, we shall derive the expressions \eqref{eq:MxD} and \eqref{eq:MxN} in the main text. Recall that we have used a uniform shifted grid scheme \cite{sarkar1989computation}, where the mesh used has a grid size of $\Delta x$ and does not coincide with the metallic boundaries; instead, relative to the boundaries, the mesh is shifted by $\Delta x/2$, as illustrated in Fig.~\ref{fig:1DMesh}. At mesh points not on the boundary, the second-order derivative can be approximated by using the central finite difference method as
\begin{align}
    \frac{d^{2}f}{dx^{2}} \approx \frac{ f_{i - 1} - 2f_i + f_{i + 1} }{\Delta x^{2}},
    \label{eq:CDiff}
\end{align}
where $f_i$ is the value of the function $f$ at point $i$.

For TE modes, the Dirichlet boundary condition is
\begin{align*}
    f_0=-f_{-1},
\end{align*}
which gives
\begin{align}
    \left.\frac{\partial^2 f}{\partial x^2}\right\vert_{x=0} & \approx
    \left.\frac{f_{i-1}-2f_i+f_{i+1}}{\Delta x^2}\right\vert_{i=0}
    \nonumber\\
    &=\frac{f_{1}-3f_{0}}{\Delta x^2}. 
    \label{eq:Eq4derive}
\end{align}

For TM modes, the Neumann boundary condition is 
\begin{align*}
    \left.\frac{\partial f}{\partial x}\right\vert_{i=0}& \approx \frac{ f_{0}-f_{-1}}{\Delta x} = 0 
    \\ 
    \Rightarrow f_0&=f_{-1},
\end{align*}
which gives 
\begin{align}
    \left.\frac{\partial^2 f}{\partial x^2}\right\vert_{x=0} & \approx
    \left.\frac{f_{i-1}-2f_i+f_{i+1}}{\Delta x^2}\right\vert_{i=0} \nonumber \\
    &=\frac{f_{1}-f_{0}}{\Delta x^2}.    
    \label{eq:Eq5derive}
\end{align}

For both the TE and TM modes, the boundary conditions at the other end can be obtained in a similar fashion to \eqref{eq:Eq4derive} and \eqref{eq:Eq5derive}, respectively. The different boundary conditions corresponding to the TE and TM modes modify the first and last entries of \eqref{eq:CDiff} to produce \eqref{eq:MxD} and \eqref{eq:MxN}, respectively.
\addedEnd

\section{Derivative of Cost Function}
\label{app:derivative}

In this appendix, we derive an analytical expression for the derivative of the cost function \eqref{eq:higher_order}. Denoting the state $\ket{\psi(\theta^{(i)})}$ by $\ket{\psi_i}$, the cost function \eqref{eq:higher_order} can be written as
\begin{align}
    F_k(\theta) &= \bra{\psi(\theta)}M \ket{\psi(\theta)} + \sum_{i=0}^{k-1} \beta_i |
    \bket{\psi(\theta)}{\psi_i}
    |^2 \nonumber\\
    &=
    \bra{\psi(\theta)} \left(
    M+\sum_{i=0}^{k-1} \beta_i \kbra{\psi_i}{\psi_i}
    \right)
    \ket{\psi(\theta)}
    \nonumber\\
    &=
    \bra{\psi(\theta)} A \ket{\psi(\theta)},
    \label{eq:costFunctionRewrite}
\end{align}
where 
\begin{align}
    A = M+\sum_{i=0}^{k-1} \beta_i \kbra{\psi_i}{\psi_i}.
\end{align}

Differentiating the cost function \eqref{eq:costFunctionRewrite} with respect to the $j$-th component of the parameter vector $\theta$ gives
\begin{align}
    \frac{\partial
    F_k(\theta)}{\partial \theta_j}  &=
    \left( \frac{\partial \bra{ \psi(\theta)}}{\partial \theta_j} \right)
    A\ket{\psi(\theta)
    }
    +
    \bra{ \psi(\theta)} A 
    \left( \frac{\partial \ket{ \psi(\theta)}}{\partial \theta_j} \right)
    \nonumber\\
    &= 2 \left( \frac{\partial \bra{ \psi(\theta)}}{\partial \theta_j} \right)
    A\ket{\psi(\theta)
    },
    \label{eq:unnormalizedDerivative}
\end{align}
where the last line follows from the hermiticity of $A$ and the fact that the hardware-efficient ansatz has a matrix representation that is real (for ansatzes that are not real, the expression \eqref{eq:unnormalizedDerivative} should be replaced by its real part). Note that the derivative in the above expression is given by
\begin{align}
    \frac{\partial \ket{ \psi(\theta)}}{\partial \theta_j} &=
    \frac{\partial}{\partial \theta_j}
    U(\theta_1, \ldots, \theta_j, \ldots ) \ket 0^{\otimes n} \nonumber\\
    &=
    \frac 12 U(\theta_1, \ldots, \theta_j + \pi,\ldots) \ket 0^{\otimes n},
\end{align}
where the last line follows from the fact that in the hardware-efficient ansatz (see Fig.~\ref{fig:ansatz}), each component $\theta_i$ of the parameter vector $\theta$ occurs exactly once in the circuit as the exponent of the rotation gate $R_y(\theta_i) = \exp(-i \theta_i Y/2)$. Since $U(\theta_1, \ldots, \theta_j + \pi,\ldots)$ is unitary, the state $\frac{\partial \ket{ \psi(\theta)}}{\partial \theta_j}$ is not normalized. We shall denote the normalized state corresponding to it as
\begin{align}
    \ket{\partial_j \psi(\theta)} := 
    2
    \frac{\partial \ket{ \psi(\theta)}}{\partial \theta_j} = U(\theta_1, \ldots, \theta_j + \pi,\ldots) \ket 0^{\otimes n}.
\end{align}
Hence, \eqref{eq:unnormalizedDerivative} may be written as
\begin{align}
    \frac{\partial F_k(\theta) }{\partial \theta_j} 
    = \bra{\partial_j \psi(\theta)}A\ket{\psi(\theta)}.
    \label{eq:normalizedDerivative}
\end{align}

Our goal is to express \eqref{eq:normalizedDerivative} as a quantum expectation value. To this end, we will make use of the following identity: if $H$ is a Hermitian $d \times d$ matrix and $\ket u$ and $\ket v$ are $d$-dimensional (complex) unit vectors, then
\begin{align}
    \Re \bra u H \ket v = \bra {u,v}X\otimes H \ket{u,v},
    \label{eq:innerProductIdentity}
\end{align}
where $X$ is the single-qubit Pauli-X operator, $\Re z = (z+\bar z)/2$ denotes the real part of a complex number z, and
\begin{align}
    \ket{u,v} &=
    \frac 1{\sqrt{2}}\big(\ket 0 \otimes \ket u + \ket 1 \otimes \ket v\big)
    = \frac{1}{\sqrt{2}}\begin{pmatrix}
    u\\v
    \end{pmatrix}.
\end{align}

Using \eqref{eq:innerProductIdentity} and the fact that the inner product in \eqref{eq:normalizedDerivative} is real, we obtain
\begin{align}
    \frac{\partial F_k(\theta)}{\partial \theta_j}  
    &=
    \bra{\partial_j \psi(\theta),\psi(\theta)}X\otimes A \ket{\partial_j \psi(\theta),\psi(\theta)}
    \nonumber\\
    &=
    \big\langle X\otimes M \big\rangle_{
    \partial_j \psi(\theta),\psi(\theta)
    }
    \nonumber\\
    &\quad + 
    \sum_{i=0}^{k-1} \beta_i
    \big\langle X\otimes \kbra{\psi_i}{\psi_i} \big\rangle_{
    \partial_j \psi(\theta),\psi(\theta)
    },
    \label{eq:gradCalc}
\end{align}
where we used the following notation to denote quantum expectation values: $   \big\langle H \big\rangle_{\phi} = \bra \phi H \ket \phi$. To express the gradient \eqref{eq:gradCalc} in terms of preparable states and simple observables, we use the fact that $\ket{\psi_i}=\ket{\psi(\theta^{(i)})} = U(\theta^{(i)}) \ket 0
$. By defining the state
\begin{align}
    \ket{\Phi_{ij}(\theta)} = \left[ I \otimes U^\dag(\theta^{(i)})\right] \ket{\partial_j \psi(\theta),\psi(\theta)},
\end{align}
the gradient \eqref{eq:gradCalc} can be written as the following sum of expectation values:
\begin{align}
    \frac{\partial F_k(\theta)}{\partial \theta_j}  
    &=
    \big\langle X\otimes M \big\rangle_{
    \partial_j \psi(\theta),\psi(\theta)
    } + 
    \sum_{i=0}^{k-1} \beta_i
    \big\langle X\otimes \kbra{0}{0} \big\rangle_{
    \Phi_{ij}(\theta)
    }.
\end{align}

\section*{Acknowledgment}

We acknowledge the use of IBM Quantum services for this work. The views expressed are those of the authors, and do not reflect the official policy or position of IBM or the IBM Quantum team.

\bibliographystyle{IEEEtran}

\vfill 
\end{document}